\documentclass[conference,10pt]{IEEEtran}
\usepackage{epsfig,epsf,rotating,setspace,latexsym,amsmath,amssymb,amsfonts,bm,theorem,subfigure,epstopdf,cite,authblk,bbm,nonfloat,comment}
\usepackage{algorithm}
\usepackage[noend]{algpseudocode}
\usepackage{color}
\usepackage{mathtools}
\usepackage{soul}

\algrenewcommand\algorithmicforall{\textbf{foreach}}
\algrenewcommand\algorithmicindent{.8em}

\newtheorem{theorem}{Theorem}
\newtheorem{lemma}{Lemma}

\newenvironment{Proof}[1]{\medskip\par\noindent{\bf Proof:\,}\,#1}{{\mbox{\,$\blacksquare$}\par}}

\IEEEoverridecommandlockouts
\allowdisplaybreaks
\begin{document}

\title{Age of Gossip With Time-Varying Topologies} 

\author{Arunabh Srivastava \qquad Thomas Jacob~Maranzatto \qquad Sennur Ulukus\\
        \normalsize Department of Electrical and Computer Engineering\\
        \normalsize University of Maryland, College Park, MD 20742\\
        \normalsize  \emph{arunabh@umd.edu} \qquad \emph{tmaran@umd.edu} \qquad \emph{ulukus@umd.edu}}

\maketitle

\begin{abstract}
We consider a gossiping network, where a source node sends updates to a network of $n$ gossiping nodes. Meanwhile, the connectivity topology of the gossiping network changes over time, among a finite number of connectivity ``states,'' such as the fully connected graph, the ring graph, the grid graph, etc. The transition of the connectivity graph among the possible options is governed by a finite state continuous time Markov chain (CTMC). When the CTMC is in a particular state, the associated graph topology of the gossiping network is in the way indicated by that state. We evaluate the impact of time-varying graph topologies on the freshness of information for nodes in the network. We use the version age of information metric to quantify the freshness of information at the nodes. Using a method similar to the first passage percolation method, we show that, if one of the states of the CTMC is the fully connected graph and the transition rates of the CTMC are constant, then the version age of a typical node in the network scales logarithmically with the number of nodes, as in the case if the network was always fully connected. That is, there is no loss in the age scaling, even if the network topology deviates from full connectivity, in this setting. We perform numerical simulations and analyze more generally how having different topologies and different CTMC rates (that might depend on the number of nodes) affect the average version age scaling of a node in the gossiping network.
\end{abstract}

\section{Introduction}\label{sec: Introduction}
Recent advances in wireless communication have led to significant improvements in information rate and reliability. The roll-out of 5G has made users more connected than ever before. These new technologies enable easy addition of and massive connectivity for new devices to the network. The enhancement in connectivity has lead to the exploration of new use cases, such as drone networks, sensor networks in remote areas, and self-driving car networks. These are usually dense networks, or may be far away from base stations. Hence, \emph{gossiping}, i.e., peer-to-peer message dissipation, has emerged as one way of disseminating information fast in such networks. Gossiping protocols are simple algorithms which involve spreading information across the entire network by communication between neighbors. Some of these applications may not have a fixed structure due to mobility or channel unreliability. Hence, there is a need to analyze how dynamic changes in graph connectivity structure impacts such networks. This analysis also helps us in the design of adaptive protocols for networks that fall under this paradigm.

Quantification of freshness of information necessitates going beyond the commonly used metrics such as latency and throughput \cite{popovski2022perspective}. Age of information was introduced as a metric to capture the freshness of information \cite{kaul2012real, sun2019age, yatesJSACsurvey}. Networks have been analyzed and designed with the age of information metric in mind \cite{yates2018age, banerjee2023re}. Several other metrics have been proposed to quantify the freshness of information in application-specific settings, including the age of incorrect information \cite{maatouk20AOII}, the age of synchronization \cite{zhong18AoSync}, binary freshness metric \cite{cho3BinaryFreshness}, and the version age of information \cite{yates21gossip, Abolhassani21version, melih2020infocom}.

\begin{figure*}
    \centering
    \includegraphics[width=0.9\linewidth]{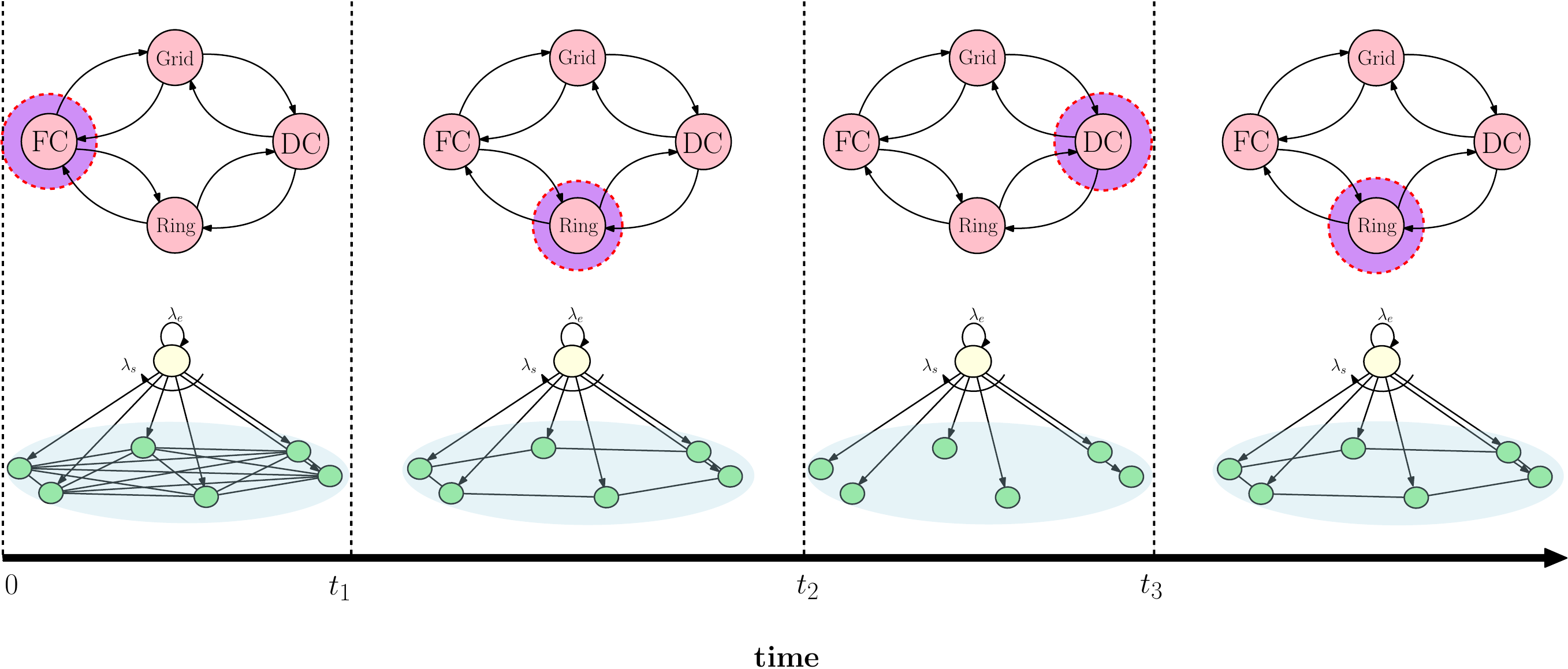}
    \caption{A description of the time evolution of the connectivity of the gossip network. The CTMC represents the connectivity graph (topology) of the gossip network. In this example, the CTMC has four states, representing the fully connected (FC) topology, the ring topology, the grid topology and the disconnected (DC) topology. The current topology is represented using a red dashed circle. In this example, the network starts in the fully connected topology, moving to the ring topology, the disconnected topology and back to the ring topology. Since the rates of the CTMC are different, the time spent in a state will be random, which is also shown in the figure.}
    \label{fig: main fig}
\end{figure*}

In this paper, we use the version age of information in a gossiping setting (age of gossip) to quantify the freshness of information. The version age of information can be defined for a node in the gossiping network as the number of versions behind the node is compared to the source node that generates the updates. These updates are typically generated by a random process. Yates \cite{yates21gossip} was the first to use the stochastic hybrid system (SHS) framework to analyze gossip networks and provide a set of recursive equations to find the version age of a subset of nodes. In particular, the version age of any subset depends on the version age of subsets containing exactly one more neighboring node. Yates showed that the long term average version age of a node in the fully connected network scales as $\Theta(\log{n})$, where $n$ is the number of nodes. 

Following Yates' work, many works have used the SHS framework to analyze gossiping networks under various settings, including community structures \cite{buyukates22ClusterGossip}, adversarial settings \cite{kaswan22jamming} and distributed computation \cite{mitra_allerton22, mitra_Infocom23}. The survey \cite{kaswan2023versionagesurvey} discusses works in this area. Recently, \cite{maranzatto2024agegossipconnectiveproperties} showed that for any gossip network, the age of a vertex (i.e., node) is equal in distribution to the first passage time from that vertex to the source. Similarly, \cite{kaswan23nonpoissonCDC} investigates how version age is affected by non-Poisson connections between nodes of the gossiping network. Many of these works have analyzed how version age scales with the number of nodes for specific topologies of the gossiping network. \cite{buyukates22ClusterGossip} shows that the version age under the ring topology scales as $\Theta(\sqrt{n})$. \cite{srivastava2024networkconnectivityinformationfreshnesstradeoff} shows that the version age scales as $O(n^{\frac{1}{3}})$ in the square grid topology, and as $O(\log{n}\log{\log{n}})$ in the unit hypercube topology. \cite{maranzatto24} analyzes complete bipartite graphs, shows that $d$-regular graphs have logarithmic version age scaling and analyzes Erdos-Renyi random graphs as well. \cite{maranzatto2024agegossipconnectiveproperties} further analyzes new graphs and gives tight upper and lower bounds for the expected version age of any vertex using a new combinatorial quantity. Gossiping networks with mobility have been analyzed using the spatial mean field regime methodology in \cite{chaintreau2009age} and SHS framework in \cite{srivastava2024mobilityagebasedgossipnetworks}. However, no paper has analyzed networks where the gossip links have time-varying rates while the source updates the gossiping network at constant rate.

In this work, we use a new method inspired by the first passage percolation methods to analyze how varying gossiping network topologies influence the version age of nodes in the network. In the system we consider, the gossiping network changes topologies according to a continuous time Markov chain (CTMC) process. These topologies are arbitrary fixed graphs, which could be a fully connected network, a ring network, a grid network or some other network. It is possible to write SHS equations for this framework in a similar way to \cite{yates2018age}. However, it is not clear if the system of equations obtained are stable and if the limiting version age exists. As an example, assume that we have two possible topologies, the fully connected topology and the ring topology, and we switch between these two states with linear rates (in the number of nodes). Then, we can show that the version age of a node moves from $\Theta(\log{n})$ to $\Theta(\sqrt{n})$ depending on the current state. Hence, instead of analyzing the limiting average version age, we analyze how the version age of a node depends on the current topology, and if the existence of certain topologies in the network ensure low or high version age for the node. In particular, we analyze a network which has constant CTMC rates, where one of the topologies is the fully connected topology. For this case, we show that the version age of a node scales as $\Theta(\log{n})$, which is the same if the network was always in a fully connected mode. That is, we show that, in this setting, there is no loss in the age scaling, even if the network topology deviates from full connectivity.

\section{System Model}\label{sec: System Model}
We consider a gossiping network consisting of a source node that generates updates and a set $\mathcal{N} = \{1,\ldots,n\}$ of $n$ gossiping nodes. The source node generates updates as a rate $\lambda_e$ Poisson process independent of all other processes in the network. The source shares its updates with node $j$ in the network as a rate $\lambda_{0j}$ Poisson process independent of all other processes. Thus, the source shares updates with the entire gossip network as a rate $\lambda_s = \sum_{j \in \mathcal{N}}\lambda_{0j}$ Poisson process.

Nodes in the gossiping network communicate with their neighbors. However, due to the changing dynamics of communication in the network, the topology with which the nodes are connected to each other changes frequently. We model this change by considering a finite set of topologies indexed by $k \in \mathcal{K} = \{1,2,\ldots,K\}$ where $K < \infty$. The network changes between these topologies as a CTMC, i.e., each individual topology exists as a state of the CTMC. The rate of leaving state $i$ is $q_i$ and the probability of moving from state $i$ to  $j$ is $p_{ij}$. We define the transition rates as $q_{ij} = q_ip_{ij}$. In topology $k$, node $i$ sends updates to neighbor $j$ according to the push-style gossip strategy as a rate $\lambda_{ij}^{(k)}$ Poisson process. We assume that the total rate with which node $i$ sends updates to its neighbors is $\lambda = \sum_{j \in N(i)}\lambda_{ij}^{(k)}$, $\forall k \in \mathcal{K}$. Our results are asymptotic in the number of nodes in the network.  We assume $K$ is constant as $n \to \infty$, but that the transitions rates and probabilities can be dependent on $n$.  We also note that the process on each node can be modeled as a doubly stochastic, Cox, process, though we do not use this characterization for the analysis.

We use the version age of information (version age) metric to quantify the freshness of information for nodes in the gossiping network. To this end, let $N_0(t)$ be the counting process associated with the version of updates at the source. Similarly, let $N_i(t)$ be the counting process associated with the version at node $i$. We note that $N_0(t)$ follows a Poisson distribution, but in general, $N_i(t)$ does not for $i \in \mathcal{N}$. We define the version age of node $i$ as $X_i(t) = N_0(t) - N_i(t)$.

The evolution of version age under the push-style gossiping protocol is as follows: If the source generates an update, then the version age of all nodes in the gossiping network increases by $1$, since each node is now one extra version behind the source. If the source sends an update to a node in the network, that node's version age drops to zero, since it now has the latest version of the update. If node $i$ sends an update to node $j$, then $j$ accepts the update if the version age of $i$ is lower than the version age of $j$, otherwise it rejects the update.

Throughout we assume $\lambda_s$, $\lambda$ are constants and let $n \rightarrow \infty$. We say that an event $\mathcal{A}$ holds with high probability(w.h.p.) if $\mathbb{P}[\mathcal{A}] \to 1$ as $n \to \infty$. We use the standard definitions for $\Theta(f(n))$, $\Omega(f(n))$, $O(f(n))$, $o(f(n))$.

\section{Effect of the Fully Connected Network Under Constant CTMC Rates}\label{sec: FC}
\begin{figure}[t]
    \centering
    \includegraphics[width=0.45\linewidth]{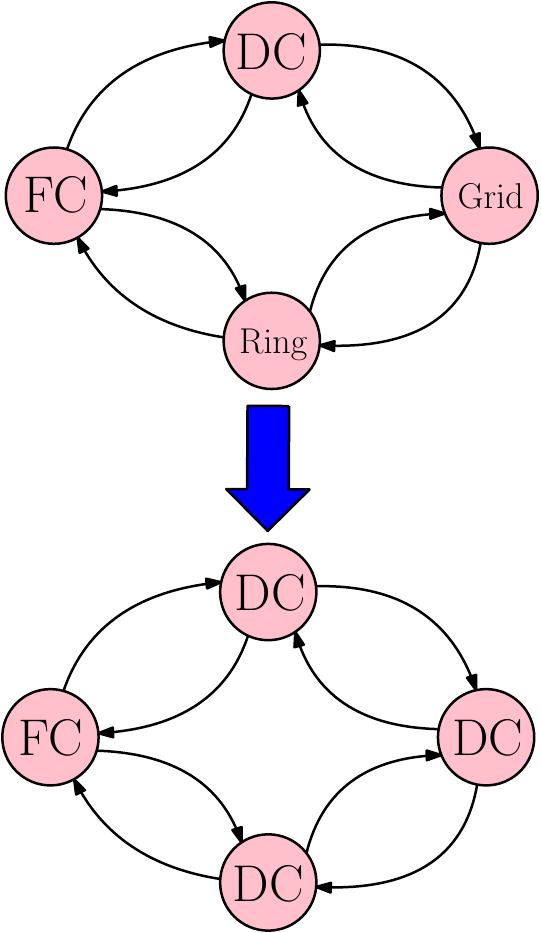}
    \caption{Top: A sample CTMC containing the fully connected (FC) topology as a state. The other states can be any fixed topology. In this example, the disconnected (DC), ring and grid topologies are present. Bottom: A network where all states are replaced by the DC topology except one FC topology. As we note in Section~\ref{sec: FC}, the average version age of a node on the top will be less than the average version age of the same node on the bottom.}
    \label{fig: fc varying topologies}
\end{figure}

In this section, we assume that each $q_i, i \in \mathcal{K}$, is constant independent of $n$. We also assume that one of the topologies is the fully connected network, where each node sends updates to every other node in the gossip network with rate $\frac{\lambda}{n-1}$. We assume that the CTMC of topologies starts in the stationary distribution. We do not make any assumption about the topology in any other state of the CTMC. This is illustrated in Fig.~\ref{fig: fc varying topologies}. In this section, we want to show that the version age of a node in the gossiping network scales logarithmically with the number of nodes in the network. Concretely, we want to show that $X_i(t) = O(\log{n})$, $\forall i \in \mathcal{N}$ w.h.p.

It was shown in~\cite{maranzatto24} that the version age does not increase with the addition of more edges in a static network. This implies that the fully connected network has the lowest version age among all possible network topologies. Since our network is not fully connected all the time, the version age performance of the network cannot be better than the performance of the fully connected network with the same rates. Hence, the lower bound for the version age of our network is $\Theta(\log{n})$. 

We use a new method to find an upper bound for the version age of a node. In order to do so, we assume that the source sends an update to node $1$ at time $0$ without loss of generality. Suppose the time it takes for this update to reach all nodes in the network is $T$ ($T$ is a stopping time). Then, the number of times the source updates itself between $0$ and $T$, given by $N_0([0,T])$, is an upper bound for the version age of each node in the network. This is because the version age of each node is how many versions behind each node is when compared to the version at the source. Hence, the version age of any node at time $T$ can be at most $N_0([0,T])$, since the source keeps updating other nodes after it has updated node $1$ at time $0$. 

For our network, we will show that $N_0([0,T]) = \Theta(\log{n})$. We formally state this in the following theorem.

\begin{theorem}\label{thm: 1}
    Suppose the source updates node $1$ at time $0$, and does not send any other updates to $\mathcal{N}$. Let $T = \inf_{t \geq 0} \ \{\forall i \in \mathcal{N}, i \text{ has received the update at time $t$} \}$. Then,
    \begin{align}
        N_0([0,T]) = \Theta(\log{n}).
    \end{align}
\end{theorem}

In order to prove this theorem, we first prove three lemmas. The first lemma shows that any update can spread from a single node to the entire fully connected network in logarithmic time. This proof follows along the lines of the proof of \cite[Thm.~1.2]{JANSON_1999}. We note that in our model, each edge has total rate $\frac{\lambda}{n-1}$, while in Janson's model in \cite{JANSON_1999} each edge has rate $1$. We include the proof below for completeness.

\begin{lemma}\label{lemma: 1}
    Suppose we are only in the fully connected topology. Then, the update sent to node $1$ at time $0$ reaches every node in the network in logarithmic time with high probability.
\end{lemma}

\begin{Proof}
    In this proof, when we say a node is updated, it means that it has received the update that was sent to node $1$ at time $0$. Node $1$ sends updates to other nodes in the network with total rate $\lambda$. Hence, a second node will get the update at a time that is exponentially distributed with rate $\lambda$. When $k<n$ nodes have been updated, then there are $k(n-k)$ links between the set of nodes that are updated and the set of nodes that are not updated. The set of updated nodes increases in size by $1$ whenever any of these $k(n-k)$ links is activated. Due to the memoryless property of the Poisson process and the fact that superposition of Poisson processes is itself a Poisson process, we see that the time it takes for the size of the updated set of nodes to increase from $k$ to $k+1$ is exponentially distributed with rate $\frac{k(n-k)\lambda}{n-1}$. 

    Next, define the time it takes for the set of nodes to increase from $k$ nodes to $k+1$ nodes as $V_k$. Note that, $V_k$ is independent of $V_{k-1}, \ldots, V_1$ due to the memoryless property of the exponential distribution. Thus, $V_1, V_2, \ldots, V_{n-1}$ are independent with $V_k \sim \text{Exponential}\left(\frac{n-1}{k(n-k)\lambda}\right)$. Let $S_m$ be the time when $m$ nodes have been updated. Then,
    \begin{align}
        S_m = \sum_{k=1}^{m-1} V_k.
    \end{align}
    The time it takes for all nodes to receive the update is the same as the time when the last node receives the update,
    \begin{align}
        S_n = \sum_{k=1}^{n-1} V_k.
    \end{align}
    We have the expectation of $S_n$ given by,
    \begin{align}
        \mathbb{E}[S_n] =& \sum_{k=1}^{n-1} \mathbb{E}[V_k]\\
        =& \sum_{k=1}^{n-1} \frac{n-1}{k(n-k)\lambda}\\
        =& \frac{n-1}{\lambda}\sum_{k=1}^{n-1} \frac{1}{k(n-k)}\\
        =& \frac{n-1}{\lambda} \sum_{k=1}^{n-1} \frac{1}{n} \left(\frac{1}{k} + \frac{1}{n-k}\right)\\
        =& \frac{2(n-1)}{n\lambda}\sum_{k=1}^{n-1} \frac{1}{k}\\
        \approx& \frac{2(n-1)}{n\lambda}(\log{n} + \gamma),
    \end{align}
    where $\gamma \approx 0.577$ is the Euler-Mascheroni constant. The variance of $S_n$ can be bounded as follows,
    \begin{align}
        \text{Var}[S_n] =& \sum_{k=1}^{n-1}\text{Var}[V_k]\\
        =& \sum_{k=1}^{n-1}\frac{(n-1)^2}{k^2(n-k)^2\lambda^2}\\
        \leq& \frac{8(n-1)^2}{\lambda^2 n^2}\sum_{k=1}^{\frac{n}{2}} \frac{1}{k^2}\\
        \leq& \frac{8(n-1)^2}{\lambda^2 n^2}\frac{\pi^2}{6}\\
        \leq& \frac{4\pi^2}{3\lambda}.
    \end{align}
    Then, using Chebyshev's inequality, we get,
    \begin{align}
        \mathbb{P}\big[|S_n - \mathbb{E}[S_n]| > f(n)\big] \leq& \frac{\text{Var}[S_n]}{(f(n))^2}\\
        =& \frac{\frac{4\pi^2}{3\lambda}}{(f(n))^2}\\
        =& o(1).
    \end{align}
    Here, $f$ is an increasing function with $f(n) = o(\log{n})$. Hence, $\frac{2(n-1)}{n\lambda}(\log{n} + \gamma) - f(n) \leq S_n \leq \frac{2(n-1)}{n\lambda}(\log{n} + \gamma) + f(n)$ with high probability.
\end{Proof} 

\begin{lemma}\label{lemma: 2}
    Assume at time $T$, the CTMC has spent $\Theta(\log n)$ time in state $k$. Then, $T = \Theta(\log{n})$ with high probability.
\end{lemma}

\begin{Proof}
    The proof follows quickly from the ergodic theorem, since $T = \Theta( \log n) \to \infty$. We give a different proof below from first principles and using absorbing Markov chains.

    For some $\beta \geq 0$, let $T$ be the first time the CTMC has spent $\beta \log n$ time in the fully connected state. Then, $T$ is a stopping time, and furthermore at time $T$, we are in the fully connected state of the CTMC. We also start in the stationary distribution, and we may start in the fully connected topology or some other state of the CTMC. If we start in the fully connected topology, we will leave the state $N_{FC}(\beta\log{n})$ times where $N_{FC}$ is the counting process associated with the Poisson process when in the fully connected topology state of the CTMC. If we start in another state of the CTMC, we will reach the fully connected state and then return to the state $N_{FC}(\beta\log{n})$ times. We note that the time to return to the fully connected state is an inter-arrival time for a renewal process whose arrival times are the instances of return to the fully connected state. These inter-arrival times, say $\{I_i\}_{i=1}^{N_{FC}(\beta\log{n})}$, follow a phase-type distribution, which is discussed in \cite{bladt2005review}.
    
    We start by finding $\text{Var}[T]$. Suppose we spend a total of $T_{out}$ time in other states up to time $T$, then $\text{Var}[T] = \text{Var}[T_{out}]$. Further, let the time of reaching the fully connected state for the first time be $T_1$, with $T_1 = 0$ if we start in the fully connected state. Then, $T_{out} = T_1 + \sum_{i=1}^{N_{FC}(\beta\log{n})}I_i$. Further, we note that $N_{FC}(\beta\log{n})+1$  is a stopping time for the Poisson process associated with the fully connected state of the CTMC. We note that $\mathbb{E}[T_1] \leq \eta \mathbb{E}[I_1]$ for some $\eta \in \mathbb{R}$. Let $\text{Var}[I_i] = \sigma^2$ and $\mathbb{E}[I_i] = \mu$. Then, we have
    \begin{align}
        \mathbb{E}[T_{out}] =& \mathbb{E}[T_1] + \mathbb{E}\bigg[\sum_{i=1}^{N_{FC}(\beta\log{n})}I_i\bigg]\\
        \leq& (\eta - 1)\mu + \mathbb{E}\bigg[\sum_{i=1}^{N_{FC}(\beta\log{n})+1}I_i\bigg]\\
        =& (\eta - 1)\mu + \mathbb{E}[N_{FC}(\beta\log{n})+1]\mu\\
        =& (\eta - 1)\mu + (q_{FC} \beta\log{n} + 1)\mu\\
        \leq& \rho \log{n},
    \end{align}
    for some $\rho \in \mathbb{R}$. The expectation is also lower bounded by the logarithm if we remove $T_1$ from the equation and write the expectation. Thus, we let $\mathbb{E}[T_{out}] = \nu \log{n}$. Due to the strong Markov property and independence of these inter-arrival times, we can further say, $\text{Var}[T_{out}] = \text{Var}[T_1] + \sum_{i=1}^{N_{FC}(\beta\log{n})}\text{Var}[I_i]$. We can further see that $\text{Var}[T_1] \leq \chi\text{Var}[I_0]$, where $\chi \in \mathbb{R}$ and $I_0$ has the same phase-type distribution as $I_1$. This is due to the positive recurrence of the CTMC. Thus, we can write $\text{Var}[T_{out}] \leq (\chi^2-1)\text{Var}[I_0] + \sum_{i=1}^{N_{FC}(\beta\log{n})+1}\text{Var}[I_i]$.

    The phase-type distribution arises from a mixture of exponential distributions, where we have multiple transient states and one absorbing state. To fit our CTMC into this framework, we divide the fully connected state of the CTMC into two states, FC1, where we start, and FC2, which is the absorbing state. Then, the variance of the inter-arrival times can be defined using the initial probability distribution of starting in a transient state, given by $\mathbf{\pi}$, and the matrix of rates of the CTMC, given by $S$. The variance is then given by,
    \begin{align}
        \text{Var}[I_i] = 2\mathbf{\pi} S^{-2} \mathbf{1} - (\mathbf{\pi} S^{-1}\mathbf{1})^2,
    \end{align}
    where $\mathbf{\pi}$ is an indicator vector of the fully connected state and $\mathbf{1}$ is the vector of all ones. We also write the expectation,
    \begin{align}
        \mathbb{E}[I_i] = -\mathbf{\pi} S^{-1}\mathbf{1}.
    \end{align}
    Since $S$ is a matrix of finite size and each element of $S$ is finite, the variance is finite. Now, we can find the upper bound for $\text{Var}[T]$,
    \begin{align}
        \text{Var}[T] =& \text{Var}[T_{out}]\\
        \leq& (\chi^2-1)\text{Var}[I_0] + \sum_{i=1}^{N_{FC}(\beta\log{n})+1}\text{Var}[I_i]\\
        =& (\chi^2-1)C + \sigma^2 \mathbb{E}[N_{FC}(\beta\log{n})+1] \notag\\
        &+ \mu \text{Var}[N_{FC}(\beta\log{n})+1]\\
        =& (\chi^2-1)C + \sigma^2 (q_{FC} (\beta\log{n})+1)\notag\\
        &+ \mu q_{FC} (\beta\log{n})\\
        \leq& \xi \log{n},
    \end{align}
    for some $\xi \in \mathbb{R}$. Then, using Chebyshev's inequality, we write
    \begin{align}
        \mathbb{P}\big[|T - \mathbb{E}(T)| > \log{n}\big] \leq& \frac{\text{Var}[T]}{(\log{n})^2}\\
        \leq& \frac{\xi}{\log{n}}\\
        =& o(1).
    \end{align}
    Hence, $\beta\log{n} + \rho\log{n} - \log{n} \leq T \leq \beta\log{n} + \rho\log{n} + \log{n}$ with high probability.
\end{Proof}

\begin{lemma}\label{lemma: 3}
    If $T = \Theta(\log{n})$, then $N_0([0,T]) = \Theta(\log{n})$ with high probability.
\end{lemma}

\begin{Proof}
    Let $T = \tau \log{n}$. We know that $N_0(t)$ is Poisson distributed with $\lambda_e t$. Then, we have that $\mathbb{E}[N_0([0,\tau \log{n}])] = \lambda_e \tau \log{n}$ and $\text{Var}[N_0([0,\tau \log{n}])] = \lambda_e \tau \log{n}$. Hence, using Chebyshev's inequality, and using $N$ as a shorthand for $N_0([0,\tau \log{n}])$, we have
    \begin{align}
        \mathbb{P}\big[|N - \mathbb{E}[N]| > \log{n}\big] \leq& \frac{\text{Var}[N]}{(\log{n})^2}\\
        =& \frac{\lambda_e \tau \log{n}}{(\log{n})^2}\\
        =& \frac{\lambda_e \tau}{\log{n}}\\
        =& o(1).
    \end{align}
    Hence, $(\lambda_e \tau-1) \log{n} \leq N_0([0,\tau \log{n}]) \leq (\lambda_e \tau+1) \log{n}$ with high probability.
\end{Proof}

Now, we use the above three lemmas to prove Theorem~\ref{thm: 1}.

\begin{Proof}
    Let $\beta \geq 0$, and let $[s_1, f_1), [s_2,f_2), \ldots,[s_k, f_k)$ be the intervals of time that the CTMC is in the fully connected state, such that $\sum_{i=1}^k (f_i - s_i) \geq \beta \log n$. Then, there exists an $\alpha$ depending only on $\beta$ so that the following holds with high probability: $f_k = \alpha \log n$, where $\alpha$ depends only on $\beta$, and $N_0[0, f_k] \leq \alpha \log n$.  The existence of such an $\alpha$ is guaranteed by Lemma~\ref{lemma: 2} and Lemma~\ref{lemma: 3}, respectively.

    Observe that the intensities of all edges in the fully connected network are the same within all intervals $[s_i, f_i)$, and all intervals are disjoint almost surely.  If we consider a gossip network only on the fully connected topology for total time $T = \sum_{i=1}^k (f_i - s_i)$, then by the observations above, stationarity gives that the distribution of arrivals on any edge is the same in the fully connected network for the interval $[0,T)$ as for the mobile network conditioned on the intervals $[s_i, f_i)$. Thus, by Lemma~\ref{lemma: 1}, with high probability, every node in the network has received the source's packet by time $f_k = \Theta(\log n)$.
\end{Proof}

\section{Numerical Results}\label{sec: Numerical Results}
In this section, we carry out simulations to examine how the topologies and CTMC rates affect the average version age of nodes in the gossiping network. For each simulation, we choose $\lambda_e = \lambda_s = \lambda = 1$.

\begin{figure}
    \centering
    \includegraphics[width=0.8\linewidth]{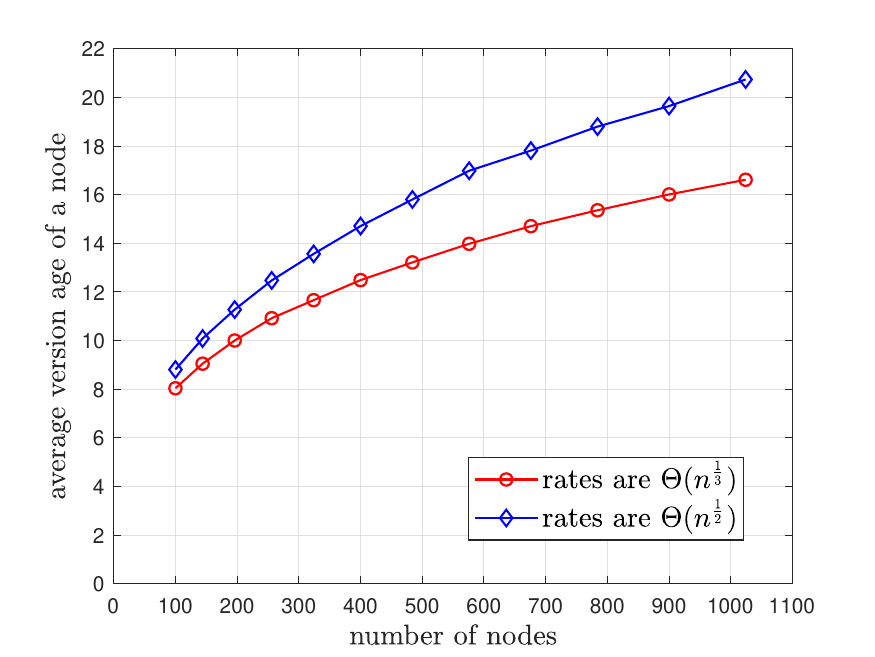}
    \caption{Variation of average version age of a node as the number of nodes varies when the CTMC has two states, the ring topology and the grid topology.}
    \label{fig: Exchange between ring and grid}
\end{figure}

First, we simulate a gossiping network where the CTMC has two states corresponding to the ring topology and the grid topology in Fig.~\ref{fig: Exchange between ring and grid}. Note that having just the ring topology will give average version age $\Theta(\sqrt{n})$ \cite{buyukates22ClusterGossip} and the grid topology will give average version age $\Theta(n^{\frac{1}{3}})$ \cite{srivastava2024networkconnectivityinformationfreshnesstradeoff, maranzatto2024agegossipconnectiveproperties}. The number of nodes varies between $100$ and $1024$ and includes all even square integers between these two. We choose two different types of rates. In the first case, we choose the rates as $\sqrt{n}$ and in the second case, we choose the rates as $n^{\frac{1}{3}}$. We observe that for the square root rates, the version age scales as $\Theta(\sqrt{n})$ as well, and for the latter case, the version age scales as $\Theta(n^{\frac{1}{3}})$.

\begin{figure}
    \centering
    \includegraphics[width=0.8\linewidth]{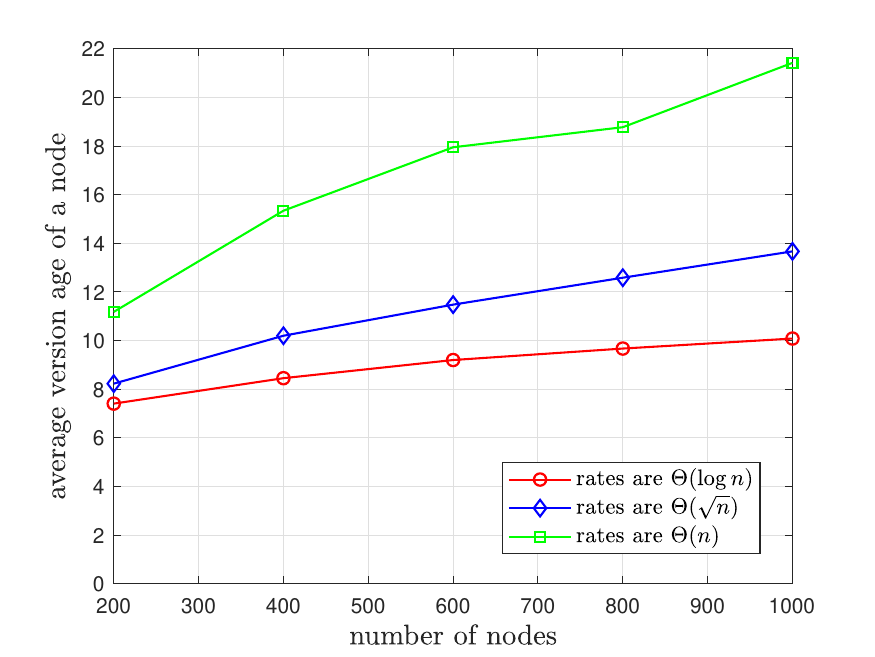}
    \caption{Variation of average version age of a node as the number of nodes varies when the CTMC has two states, the ring topology and the fully connected topology.}
    \label{fig: Exchange between ring and fc}
\end{figure}

Next, we simulate the gossiping network which has two states corresponding to the ring topology and the fully connected topology in Fig.~\ref{fig: Exchange between ring and fc}. The number of nodes varies between $200$ and $1000$ with steps of $200$. We choose three different types of rates in this case. We choose linear rates $n$, square root rates $\sqrt{n}$, and logarithmic rates $\log{n}$. The rates are the same for both states. We observe that the average version age grows slowly when the rates are logarithmic, but grows faster when the rates are square root and linear. We observe that the version age scaling is $\Theta(\sqrt{n})$ when the rates are linear and square root, and $\Theta(\log{n})$ when the rates are logarithmic.

\begin{figure}
    \centering
    \includegraphics[width=0.8\linewidth]{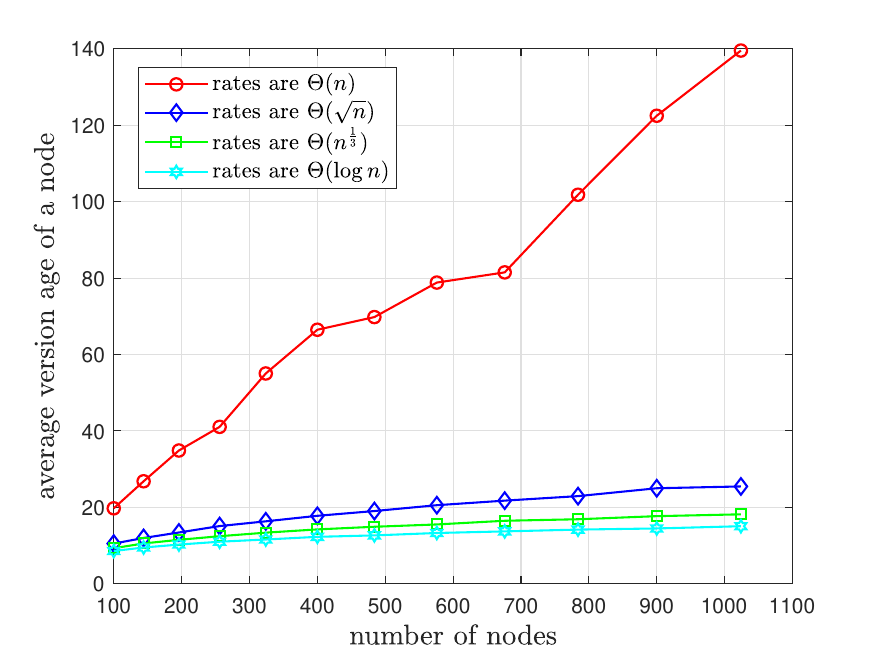}
    \caption{Variation of average version age of a node as the number of nodes varies when the CTMC has four states, the ring topology, the grid topology, the disconnected topology and the fully connected topology.}
    \label{fig: disconnected simulation}
\end{figure}

Finally, we simulate the network where the CTMC is the upper CTMC in Fig.~\ref{fig: fc varying topologies} and plot the average version age in Fig.~\ref{fig: disconnected simulation}. We note that the average version age is linear if only the disconnected topology were present. We choose four different types of rates, $n$, $\sqrt{n}$, $n^{\frac{1}{3}}$ and $\log{n}$, for each state. We assume the probability of moving to both neighboring states is equal to half. The number of nodes varies between $100$ and $1024$ and includes all even square integers between these two. We observe that the version age scales linearly if the rates are linear, and the version age as a function of $n$ gradually becomes smaller as the rates become smaller. We observe that the version age scaling is the same as the rate (among the four rates given) in this network.

\section{Conclusion}\label{sec: Conclusion}
We investigated the effects of time-varying network topologies on the version age in gossip networks. We defined a system in which the topology of a network changes according to the states of a CTMC. We considered a system which contained the fully connected topology with constant CTMC rates and showed that the version age scaling is logarithmic irrespective of the constant fraction of time spent in the fully connected topology. We carried out numerical simulations and observed that the average version age of a node depends on the topologies that are possible and the rates with which the states of the CTMC change. 

\bibliographystyle{unsrt}
\bibliography{refs}

\end{document}